\documentclass[12pt]{article}

\usepackage[margin=1in]{geometry}
\usepackage[numbers,sort&compress]{natbib} 
\usepackage{authblk} 
\usepackage{color}
\usepackage{amsmath}
\usepackage{amssymb} 
\usepackage{graphicx} 
\usepackage{mathrsfs} 
\usepackage[format=plain,labelformat=simple]{subcaption}

\usepackage{fancyhdr} 
\usepackage{amsfonts}
\usepackage{tabularx} 
\usepackage{multirow}
\usepackage{slashed}
\usepackage{xcolor}
\usepackage{booktabs} 
\usepackage{bbm}
\usepackage{nicefrac} 
\usepackage[
singlelinecheck=false 
]{caption}
\usepackage{feynmp} 
\usepackage[bookmarks=false,hidelinks]{hyperref}

\usepackage[utf8]{inputenc}
\def\be{\begin{equation}}
\def\ee{\end{equation}}
\def\bea{\begin{eqnarray}}
\def\eea{\end{eqnarray}}

\newcommand{\Figref}[1]{Fig.~\ref{#1}}
\newcommand{\Tabref}[1]{Tab.~\ref{#1}}

\newcommand{\Appref}[1]{App.~\ref{#1}}

\newcommand*{\rep}[2][]{\ensuremath{{\boldsymbol{#2}#1}}}

\newcommand{\bs}[1]{\ensuremath{\boldsymbol{#1}}}
\newcommand{\I}{\mathrm{i}}

\newcommand{\U}[1]{\ensuremath{\mathrm{U}(#1)}}
\newcommand{\D}[1]{\ensuremath{\mathrm{D}_{#1}}}
\newcommand{\SU}[1]{\ensuremath{\mathrm{SU}(#1)}}

\definecolor{darkgreen}{HTML}{109930}

\begin{document}

\renewcommand\headrule{} 

\title{{\bf{A ``Vector-like chiral'' fourth family to explain muon anomalies}}}
\author{Stuart Raby$^1$\footnote{raby.1@osu.edu} }
\author{Andreas Trautner$^2$\footnote{atrautner@uni-bonn.de}}
\affil{$^1$\emph{Department of Physics}\\\emph{The Ohio State University}\\\emph{191 W.~Woodruff Ave, Columbus, OH 43210, USA}}
\affil{$^2$\emph{Bethe Center for Theoretical Physics and} \\ \emph{Physikalisches Institut der Universit\"at Bonn,}\\
\emph{Nussallee 12, 53115 Bonn, Germany}}
\maketitle
\thispagestyle{empty}
\pagenumbering{gobble} 

\begin{abstract}\normalsize\parindent 0pt\parskip 5pt
The Standard Model (SM) is amended by one generation of quarks and leptons which are vector-like (VL) under the SM gauge group but chiral with respect to a new $\U1_{3-4}$ gauge symmetry.
We show that this model can simultaneously explain the deviation of the muon $g-2$ as well as the observed anomalies in $b\rightarrow s\mu^+\mu^-$ transitions
without conflicting with the data on Higgs decays, lepton flavor violation, or $B_s-\bar{B}_s$ mixing.
The model is string theory motivated and GUT compatible, i.e.\ UV complete,
and fits the data predicting VL quarks, leptons and a massive $Z'$ at the $\mathrm{TeV}$ scale,
as well as $\tau\to3\mu$ and $\tau\to\mu\gamma$ within reach of future experiments.
The Higgs couplings to SM generations are automatically aligned in flavor space.
\end{abstract}

\pagenumbering{arabic} 

\newpage
\section{INTRODUCTION}
The Standard Model (SM) is a highly successful theory in predicting and fitting many experimental measurements, with few exceptions.
One of the discrepancies between the SM prediction and experimental measurement that has been known for a long time, is the muon anomalous magnetic moment.
The discrepancy between the measured value and the SM prediction is \cite{Bennett:2006fi, Olive:2016xmw}
\begin{align}\label{eq:anomalousg}
  \Delta a_\mu
  = a_\mu^\text{exp}-a_\mu^\text{SM}
  = 288(63)(49)\times10^{-11} \,.
\end{align}
More recently, there appeared deviations from the SM predictions in $b\rightarrow s\mu^+\mu^-$ transitions related to
tests of lepton flavor universality in the observables $R(K)$ and $R(K^*)$ \cite{Aaij:2014pli, Aaij:2014ora}, semi-leptonic branching ratios
\cite{Aaij:2013aln, Lees:2013nxa, Aaij:2015esa},
and angular distributions \cite{Aaij:2013qta, Aaij:2015oid, CMS:2017ivg, Khachatryan:2015isa, Abdesselam:2016llu, Wehle:2016yoi,ATLAS:2017dlm}.
Most interestingly, all of the more recent anomalies can simultaneously be explained 
\cite{Altmannshofer:2017fio, Altmannshofer:2017yso, Alok:2017sui, Capdevila:2017bsm, Ciuchini:2017mik, DAmico:2017mtc, Geng:2017svp, Ghosh:2017ber, Hiller:2017bzc}
by specific deviations from the SM in one or more of the Wilson coefficients $C_{9}$, $C'_{9}$, $C_{10}$ and/or $C'_{10}$ of the
effective Hamiltonian \cite{Buras:1994dj,Bobeth:1999mk} (see \cite{Bardhan:2017xcc} for a possible role of additional tensor operators)
\begin{align}
  \mathcal H_\text{eff}= -\frac{4\,G_\mathrm{F}}{\sqrt2}\,V_{tb}V_{ts}^*\,\frac{e^2}{16\pi^2}\sum_{j=9,10}(C_{j}\mathcal{O}_j + C'_{j}\mathcal{O}'_j)+\mathrm{h.c.}\;,
\end{align}
where
\begin{align} \label{eq:Wilson}
  \mathcal O_9 &= \left(\bar{s}\gamma_\mu P_L b\right)\left(\bar{\mu}\gamma^\mu \mu\right)\;,&
  \mathcal O'_9 &= \left(\bar{s}\gamma_\mu P_R b\right)\left(\bar{\mu}\gamma^\mu\mu\right)\;,&  \\
  \mathcal O_{10} &= \left(\bar{s}\gamma_\mu P_L b\right)\left(\bar{\mu}\gamma^\mu \gamma_5  \mu\right)\;, &
  \mathcal O'_{10} &= \left(\bar{s}\gamma_\mu P_R b\right)\left(\bar{\mu} \gamma^\mu \gamma_5  \mu\right)\;.&
\end{align}
A simple extension of the SM that can explain the discrepancy of the muon $g-2$ are VL leptons that couple exclusively to muons \cite{Czarnecki:2001pv, Kannike:2011ng,Dermisek:2013gta}.
On the other hand, the anomalies in $b\rightarrow s\mu^+\mu^-$ transitions can be explained by a new massive vector boson of a spontaneously broken $U(1)_{\mu-\tau}$ gauge symmetry
and the introduction of VL quarks \cite{Altmannshofer:2014cfa,Crivellin:2015mga} (see \cite{King:2017anf} for a generalization of the new gauge symmetry).
Indeed, it has been shown that combining an additional $Z'$, VL leptons, and VL quarks
one can successfully address both the muon $g-2$ and the anomalous $B$ physics observables simultaneously \cite{Allanach:2015gkd, Altmannshofer:2016oaq, Megias:2017dzd}.
Typically these models predict significant deviations of the SM in $h\rightarrow\mu\mu$ \cite{Kannike:2011ng,Dermisek:2013gta}, $h\rightarrow\mu\tau$ \cite{Crivellin:2015mga,Altmannshofer:2016oaq}
and have an upper bound on the $Z'$ mass by keeping $B_s-\bar{B}_s$ oscillations close to their SM value \cite{Altmannshofer:2014cfa}.

In the present paper we suggest a holistic way of solving the discrepancies in $(g-2)_\mu$ and $b\rightarrow s\mu^+\mu^-$.
We amend the SM by one complete family of fermions, i.e.\ a full spinor representation of $SO(10)$, which is VL with respect to the SM but chiral with respect to a new
spontaneously broken ``$U(1)_{3-4}$'' gauge symmetry. Under the new gauge group the third SM family and the left-handed part of the new ``VL'' family have
charges $+1$ and $-1$, respectively, while all other fermions are neutral. Our model is motivated by heterotic string orbifold constructions \cite{Kobayashi:2004ud,Kobayashi:2004ya,Buchmuller:2005jr,Lebedev:2006kn,Lebedev:2007hv,Lebedev:2008un,Blaszczyk:2009in,Kappl:2010yu},
which, in addition to the full MSSM spectrum, typically contain myriads of states which are VL with respect to the SM gauge group, but chiral under new $U(1)'$ gauge symmetries.
In addition, there are many SM singlet scalars that break the additional gauge symmetries, thus giving mass to the vector-like states and the extra gauge bosons.
While in earlier constructions these extra states were lifted to the string scale, our model is a prototype of what can happen if at least one extra generation is kept light,
i.e.\ at the $\mathrm{TeV}$ scale. Our analysis is not supersymmetric, but it could easily be extended to a supersymmetric model in which case gauge coupling
unification is maintained.

We find that the model can simultaneously fit the observed quark and lepton masses, as well as the $(g-2)_\mu$ and $b\rightarrow s\mu^+\mu^-$ anomalies without violating
bounds from electroweak precision observables, lepton flavor violating (LFV) decays or $B_s-\bar{B}_s$ mixing.
Interestingly, the electroweak singlet Higgs boson couplings in our model are automatically aligned with the SM values to a very high degree.
Contrary to \cite{Altmannshofer:2014cfa,Crivellin:2015mga,Altmannshofer:2016oaq} there is no upper bound on the $Z'$ mass from the $B_s-\bar{B}_s$ mixing constraint,
simply because the ``VL'' fermions and the $Z'$ simultaneously obtain mass of the order of the $U(1)_{3-4}$ breaking scale.

To substantiate our arguments we present two data points that can fit all measured observables
while predicting others. The masses of new quarks and leptons, as well as of the new $Z'$ are all at the $\mathrm{TeV}$ scale.
The $Z'$ in our example has very suppressed couplings to the first family, meaning that $Z'$ production at the LHC is suppressed.
$B_s-\bar{B}_s$ mixing is predicted to deviate from the SM at the level of a few percent. There are significant enhancements
in $\mathrm{BR}(B_s\to K^{(*)}\tau\bar{\tau})$ and $\mathrm{BR}(B_s\to \phi\tau\bar{\tau})$, while $R_{K^{(*)}}^{\nu\bar{\nu}}$ is suppressed.
Furthermore, our best fit points predict $\mathrm{BR}(\tau\to\mu\gamma)$ and $\mathrm{BR}(\tau\to3\mu)$ in reach of upcoming experiments.

\section{MODEL}
\label{sec:ch2.model}
\begin{table}[!htbp]
  \begin{center}
    \begin{tabular}{lccccccc}
      \toprule
      $G_\mathrm{SM}$ family & $\left(\rep{3},\rep{2}\right)_{\frac16}$ & $\left(\rep{3},\rep{1}\right)_{\frac23}$ & $\left(\rep{3},\rep{1}\right)_{-\frac13}$ & $\left(\rep{1},\rep{2}\right)_{-\frac12}$ & $\left(\rep{1},\rep{1}\right)_{-1}$ & $\left(\rep{1},\rep{1}\right)_{0}$ & $\U1_{3-4}$ \\
     \midrule
      $a=1,2$ &$q_\mathrm{L}^{a}=\left(u^a_\mathrm{L},d^a_\mathrm{L}\right)$ & $u_\mathrm{R}^a$ & $d_\mathrm{R}^a$ & $\ell_\mathrm{L}^a=\left(\nu^a_\mathrm{L},e^a_\mathrm{L}\right)$ & $e_\mathrm{R}^a$ & $\nu_\mathrm{R}^a$ & $0$ \\
      $3$ & $q_\mathrm{L}^3=\left(u^3_\mathrm{L},d^3_\mathrm{L}\right)$ & $u_\mathrm{R}^3$ & $d_\mathrm{R}^3$ & $\ell_\mathrm{L}^3=\left(\nu^3_\mathrm{L},e^3_\mathrm{L}\right)$ & $e_\mathrm{R}^3$ & $\nu_\mathrm{R}^3$ & $1$ \\
        $4_\mathrm{L}$ & \hspace{1pt}$Q_\mathrm{L}=\left(U'_\mathrm{L},D'_\mathrm{L}\right)$ & $U_\mathrm{R}$ & $D_\mathrm{R}$ & \;\hspace{1pt}$L_\mathrm{L}=\left(N'_\mathrm{L},E'_\mathrm{L}\right)$ & $E_\mathrm{R}$ & $N_\mathrm{R}$ & $-1\phantom{-}$ \\
      $4_\mathrm{R}$ & \,$Q_\mathrm{R}=\left(U'_\mathrm{R},D'_\mathrm{R}\right)$ & $U_\mathrm{L}$ & $D_\mathrm{L}$ & \;\,$L_\mathrm{R}=\left(N'_\mathrm{R},E'_\mathrm{R}\right)$ & $E_\mathrm{L}$ & $N_\mathrm{L}$ & $0$\\
      \bottomrule
    \end{tabular}
    \\[0.3cm]
    \begin{tabular}{lccc}
      \toprule
      & $H$ & $\Phi$ & $\left(\varphi_1,\varphi_2\right)$ \\
      \midrule
      $G_\mathrm{SM}$ & $\left(\rep{1},\rep{2}\right)_{\frac12}$ & $\left(\rep{1},\rep{1}\right)_{0}$ & $\left(\rep{1},\rep{1}\right)_{0}$ \\
      $\U1_{3-4}$ & 0 & 1 & 0 \\
      $\mathrm{D}_4$ & \rep{1} & \rep{1} & \rep{2} \\
      \bottomrule
    \end{tabular}
  \end{center}
  \caption{The quantum numbers of fermions and scalars in our model under the SM gauge group and under the new $\U1'\equiv\U1_{3-4}$.
  Note that primed fields have nothing to do with the $\U1'$ \textit{per se} but are used to denote constituents of $\SU2_\mathrm{L}$ doublets.
  }\label{tab:SM4}
  \label{tab:qm}
\end{table}
The model under investigation is the SM with three right-handed neutrinos extended by a complete extra generation of left-chiral fields and a complete extra generation of right-chiral fields.
Furthermore, we introduce a new ``$U(1)_{3-4}$'' gauge symmetry under which the third SM generation as well as the left-chiral part of the fourth generation of particles is charged.
The $U(1)_{3-4}$ gauge symmetry is spontaneously broken by the vacuum expectation value (VEV) of the new scalar $\Phi$.
All fields and their corresponding quantum numbers are summarized in Table \ref{tab:qm}. The relevant part of the Lagrangian for this study is given by
\begin{equation}
 \mathcal{L}\supset\mathcal{L}_{3,H}+\mathcal{L}_{\mathrm{VL},H}+\mathcal{L}_{3,\Phi}+\mathcal{L}_{\mathrm{VL},\Phi}+\mathcal{L}_{12,\varphi}+\mathcal{L}_\mathrm{Maj}\;,
\end{equation}
with\footnote{%
In our notation $P_L N_L=N_L$ is a two component Dirac spinor which can be written
in terms of the two component Weyl spinor, $\eta$,  as $N_L =(\eta,0)^{\mathrm{T}}$.
Then $N_L^{\mathcal{C}}=(0,\I\sigma_2\eta^*)^\mathrm{T}$.
}
\begin{align}
 \mathcal{L}_{3,H}:=& 
 -y_b\,\bar{q}_L^3H d_R^3 - y_\tau\,{\bar{\ell}_{L}}^3 H e_{R}^3 - y_\nu\,{\bar{\ell}_{L}}^3 \widetilde{H} \nu_{R}^3 +\mathrm{h.c.}\;,\\
 \mathcal{L}_{3,\Phi}:=&-\lambda_3\,\Phi \left(\bar{q}_L^3 Q_R + 
                         \bar{d}_R^3 D_L +  \bar{\ell}_L^3 L_R + \bar{e}_R^3 E_L \right)+\mathrm{h.c.}\;,  \\
 \mathcal{L}_{\mathrm{VL},H}:=&-\lambda_{LR}\left(
                                 \bar{Q}_L H D_R + \bar{L}_L H E_R + \bar{L}_L \widetilde{H} N_R\right)+\mathrm{h.c.} \\
                              &-\lambda_{RL}\left(
                                 \bar{Q}_R H D_L + \bar{L}_R H E_L + \bar{L}_R \widetilde{H} N_L \right)+\mathrm{h.c.}\;,  \\
 \mathcal{L}_{\mathrm{VL},\Phi}:=& -\Phi^*\left( \lambda_Q \bar{Q}_L Q_R 
 + \lambda_D \bar{D}_R D_L + \lambda_L \bar{L}_L L_R + \lambda_E \bar{E}_R E_L + \lambda_N \bar{N}_R N_L \right)+\mathrm{h.c.}\;, \\
 \mathcal{L}_{12,\varphi}:=&-\lambda_2\,\varphi^a \left(\bar{q}_L^a Q_R + 
                         \bar{d}_R^a D_L +  \bar{\ell}_L^a L_R + \bar{e}_R^a E_L \right)+\mathrm{h.c.}\;.  \\
 \mathcal{L}_{\mathrm{Maj}}:=&-\frac{1}{2}M_L \overline{N_L^\mathcal{C}} N_L -\frac{1}{2} M_R^{ab}\overline{\left(\nu_R^{a}\right)^\mathcal{C}} \nu_R^b - \left(M_R \overline{N_R^\mathcal{C}} \nu_R^3+\mathrm{h.c.}\right)\;.
\end{align}
We take all couplings to be real and -- in some GUT spirit -- set many of them alike.
Couplings to the up quark sector electroweak singlets ($u^a_R$, $u^3_R$, $U_R$, and $U_L$) are not displayed because they will not be constrained by our analysis. 
It is summed over the repeated indices $a=1,2$ of the $\rep{2}\oplus\rep{1}$ flavor structure of the SM families which can, for example,
originate from a $\D4$ flavor symmetry \cite{Kobayashi:2004ya,Kobayashi:2006wq,Lebedev:2007hv,Ko:2007dz,Blaszczyk:2009in,Kappl:2010yu}.
The first and second families are distinguished by the direction of the $\D4$
breaking VEV $\langle\varphi_a\rangle=\delta_{a2} v_\varphi$.
We assume this alignment to happen at a high-scale $M$ (one should imagine $M \sim M_{string}$ or $M \sim M_{\mathrm{GUT}}$), and the corresponding effective operator coefficient, thus,
should be imagined as $v_\varphi\equiv \langle\tilde{\Phi}\rangle \langle\varphi_2\rangle/M$ where $\tilde \Phi$ is a SM and $\U1_{3-4}$ neutral scalar that gets a VEV around the weak scale.
This justifies our assumption here that the first family does not directly mix with the VL states. We will focus on the
flavor structure of the second and third generations in this study,
remarking that the first family can always be fit in.
A more detailed analysis should include all three families and their flavor physics, but that is beyond the scope of the present paper.

\subsubsection*{Charged Lepton and Down Quark Masses}
The charged lepton mass terms are given by
\begin{equation}  \label{eq:leptonmass}
  \bar{\bs{e}}_{L}^A \ \mathcal{M}^\ell_{AB} \ \bs{e}_{R}^B\equiv
  \begin{pmatrix} \bar{E}_L \\ \bar{E}'_L \\ \bar e_{L}^3 \\ \bar e_{L}^2 \end{pmatrix}^{\hspace{-0.2cm}\mathrm{T}}
  \begin{pmatrix}
    \lambda_{RL}\,v & \lambda_E\,v_\Phi &   \lambda_3\,v_\Phi & \lambda_2\,v_\varphi \\
    \lambda_L\,v_\Phi & \lambda_{LR}\,v &  0 & 0 \\
    \lambda_3 \,v_\Phi &  0 &  y_\tau\,v & 0 \\
   \lambda_2\,v_\varphi &  0 &  0 &  y_\mu\,v \\
  \end{pmatrix}
  \begin{pmatrix} E'_R \\ E_R \\  e_{R}^3 \\ e_R^2 \end{pmatrix} \;,
\end{equation}
where $A,B=1,..,4$ and the scalar VEVs and couplings are all assumed to be real.
Analogously, the down quark mass terms are given by
\begin{equation}
 \bar{\bs{d}}_{L}^A \ \mathcal{M}^d_{AB} \ \bs{d}_{R}^B \;,
\end{equation}
with
\begin{align}
 \bs{d}_{L}&:= \left( D_L,\, D'_L,\, d^3_L,\, d^2_L \right) \\ \label{eq:downstates}
 \bs{d}_{R}&:= \left( D'_R,\, D_R,\, d^3_R,\, d^2_R \right) \;.
\end{align}
The matrix $\mathcal{M}^d$ has exactly the same structure as $\mathcal{M}^\ell$ with the replacements $\lambda_E\rightarrow\lambda_D$, $\lambda_L\rightarrow\lambda_Q$,
$y_\tau\rightarrow y_b$, and $y_\mu\rightarrow y_s$.
Let $U^{\ell,d}_L$ and $U^{\ell,d}_R$ be unitary matrices that diagonalize the respective mass matrix,
\begin{align}
  (U^\ell_L)^\dagger \, \mathcal{M}^\ell \, U^\ell_R&=\left(\mathcal{M}^\ell\right)^{\text{diag}}\equiv\mathrm{diag}\left( m_{E}, m_{L}, m_{\tau}, m_\mu\right)\;,\\
  (U^d_L)^\dagger \, \mathcal{M}^d \, U^d_R&=\left(\mathcal{M}^d\right)^{\text{diag}}\equiv\mathrm{diag}\left( m_{D}, m_{Q}, m_{b}, m_s\right)\;.
\end{align}
The physical fields in the mass basis are then given by
\begin{align}
 \left[\hat{\bs{e}}_{L,R}\right]^A =\left[(U^\ell_{L,R})^\dagger\right]^{A B} \left[\bs{e}_{L,R}\right]^{B} \quad\text{and}\quad \left[\hat{\bs{d}}_{L,R}\right]^A =\left[(U^d_{L,R})^\dagger\right]^{A B} \left[\bs{d}_{L,R}\right]^{B}\;.
\end{align}

\subsubsection*{Neutrino Masses}
Defining the vectors
\begin{align}
 \bs{\nu}_{L}:= \left( N_L,\, N'_L,\, \nu^3_L,\, \nu^2_L \right) \quad\text{and}\quad  \bs{\nu}_{R}&:= \left( N'_R,\, N_R,\, \nu^3_R,\, \nu^2_R \right) \;,
\end{align}
the neutrino masses can be written as
\begin{equation}\label{eq:nuMasses}
 \begin{pmatrix} \bar{\bs{\nu}}_L \\ \bar{\bs{\nu}}^\mathcal{C}_R\end{pmatrix}^{\hspace{-0.2cm}\mathrm{T}}
 \begin{pmatrix} \mathcal{M}_L & \mathcal{M}_D \\ \mathcal{M}^\mathrm{T}_D & \mathcal{M}_R \end{pmatrix}
 \begin{pmatrix} {\bs{\nu}}^\mathcal{C}_L \\ {\bs{\nu}}_R \end{pmatrix}\equiv
 \bar{\bf{N}}_L^\alpha\,\mathcal{M^\nu}_{\alpha\beta}\,{\bf{N}}_R^\beta\;,
\end{equation}
where $\alpha,\beta=1,..,8$ and ${\bf{N}}_L^\mathcal{C}={\bf{N}}_R$.
The Dirac mass terms $\mathcal{M}_D$ have the same structure as $\mathcal{M}^\ell$ with the replacements
$\lambda_E\rightarrow\lambda_N$, $y_\tau\rightarrow y_{\nu_1}$, and $y_\mu\rightarrow y_{\nu_2}$. The Majorana
mass terms $\mathcal{M}_{L,R}$ have non-zero elements $2\left[\mathcal{M}_{L}\right]_{11}= M_L$,
$\left[\mathcal{M}_{R}\right]_{23}=\left[\mathcal{M}_{R}\right]_{32}=M_R$ and $2 \left[\mathcal{M}_{R}\right]_{44}= M_R^{11}$
with all other elements being zero. Assuming the hierarchy $M\sim M_{L,R}\gg v_{\Phi,\varphi}\gg v$ the neutrino mass matrix can be analytically diagonalized
and we give details about that in \Appref{app:NeutrinoMasses}. The physical states are
\begin{equation}
 \hat{\bf{N}}^\alpha_L =\left[(U^\nu)^\mathrm{T}\right]^{\alpha\beta} {\bf{N}}^{\beta}_L\;,\qquad\hat{\bf{N}}^\alpha_R =\left[(U^\nu)^\dagger\right]^{\alpha\beta} {\bf{N}}^{\beta}_R\;,
\end{equation}
with corresponding masses
\begin{equation}
 (U^\nu)^\mathrm{T} \, \mathcal{M}^\nu \, U^\nu =\left(\mathcal{M}^\nu\right)^{\text{diag}}\approx\mathrm{diag}\left( M, M, \frac M2, \frac M2, M_D, M_D, 0 , 0  \right)\;,
\end{equation}
up to corrections of the order $v_{(\Phi,\varphi)}^2/M$.
There are four sterile neutrinos with mass at the high scale.
Furthermore, there are two light active neutrinos with mass of order $M_W^2/M_{GUT}$ and one (mostly) Dirac neutrino with a $\mathrm{TeV}$ scale mass (cf.\ \Appref{app:NeutrinoMasses})
\begin{equation}
M_D=\sqrt{\left(\lambda_L\,v_\Phi\right)^2 + \left(\lambda_3\,v_\Phi\right)^2 + \left(\lambda_2\,v_\varphi\right)^2}\;.
\end{equation}
Adding the first generation back in gives one additional high scale sterile neutrino and one additional light active neutrino.

\subsubsection*{$\boldsymbol{Z}$-Lepton Couplings}
The $Z$-lepton couplings in the mass basis are
\begin{align}
  \mathcal{L} \supset Z_\mu\left(\hat{\bar{\bs{e}}}_{L}^A\, \gamma^\mu \left[\hat{g}_{L}^Z\right]_{A B}\, \hat{\bs{e}}_{L}^B + \hat{\bar{\bs{e}}}_{R}^A \, \gamma^\mu \left[\hat{g}_{R}^Z\right]_{A B}\,\hat{\bs{e}}_{R}^B\right)\;,
\end{align}
with coupling matrices
\begin{equation}
\hat g_{L,R}^Z = (U^\ell_{L,R})^\dagger g_{L,R}^Z U^\ell_{L,R}\;.
\end{equation}
The un-hatted coupling matrices are in the gauge basis and given by
\begin{equation}
 g^Z_{L,R}=\frac{g}{c_W}\left[\mathbbm{1}\,g^{Z,\mathrm{SM}}_{L,R}\pm\mathrm{diag}\left(\frac12,0,0,0\right)\right]\;,
\end{equation}
where $g^{Z,\mathrm{SM}}_{L}=(-1/2+s_W^2)$, $g^{Z,\mathrm{SM}}_{R}=s_W^2$, and we use the abbreviations $s_W=\sin\theta_W$, $c_W=\cos\theta_W$.
Since these matrices are not proportional to the identity matrix, the $Z$-lepton couplings are not diagonal in the mass basis.
Hence, this model has LFV $Z$ boson decays, which are, however, only effective amongst the heavy VL quarks and leptons.

\subsubsection*{$\boldsymbol{W}$-Lepton Couplings}
The $W$-lepton couplings in the mass basis are
\begin{equation}\label{eq:W}
  \mathcal L \supset W_\mu^+\left(\hat{\bar{\mathbf{N}}}_L^\alpha \gamma^\mu \left[\hat{g}^W_L\right]_{\alpha B} \hat{{\bs{e}}}_{L}^B +
  \hat{\bar{\mathbf{N}}}^\alpha_R \gamma^\mu \left[\hat{g}^W_R\right]_{\alpha B} \hat{\bs{e}}_{R}^B\right) + \text{h.c.} \;,
\end{equation}
where
\begin{align}
 \hat g_{L}^W &= \frac{g}{\sqrt{2}}[(U^\nu)^\mathrm{T}g^W_L U^\ell_L]& &\text{and}& \hat g_{R}^W& = \frac{g}{\sqrt{2}}[(U^\nu)^\dagger g^W_R U^\ell_R]\;,&
\end{align}
with the $8\times4$ coupling matrices of the gauge basis
\begin{equation}
 g^W_L=\begin{pmatrix} \mathrm{diag}\left(0,1,1,1\right)\\ \boldsymbol{0}_{4\times4} \end{pmatrix} \quad\text{and}\quad
 g^W_R=\begin{pmatrix} \boldsymbol{0}_{4\times4} \\ \mathrm{diag}\left(1,0,0,0\right)\end{pmatrix}\;.
\end{equation}

\subsubsection*{$\boldsymbol{Z'}$ Couplings}
The $Z'$ couplings to charged leptons in the mass basis are
\begin{equation}\label{eq:ZpLagrangian}
  \mathcal L \supset g' Z'_\mu\left(\hat{\bar{\bs{e}}}_{L}^A \gamma^\mu \left[\hat{g}^\ell_L\right]_{AB} \hat{\bs{e}}_{L}^B +
  \hat{\bar{\bs{e}}}_{R}^A \gamma^\mu \left[\hat{g}^\ell_R\right]_{AB} \hat{\bs{e}}_{R}^B \right) \;,
\end{equation}
where
\begin{equation}
 \hat{g}^\ell_{L,R} = (U^\ell_{L,R})^\dagger g_{L,R} U^\ell_{L,R}\;,
\end{equation}
with the $\U1_{3-4}$ charge matrices
\begin{equation}
 g_{L}=g_{R}=\mathrm{diag}(0,-1,1,0)\;.
\end{equation}
The  $Z'$ couplings here are \emph{not} left-right symmetric, recall the charge assignment \Tabref{tab:SM4},
and our skewed definition of the right-handed states in \eqref{eq:leptonmass} and \eqref{eq:downstates}.
The $Z'$-down quark couplings in the mass basis are completely analogously given by
\begin{equation}
  \mathcal L \supset g' Z'_\mu\left(\hat{\bar{\bs{d}}}_{L}^A \gamma^\mu \left[\hat{g}^d_L\right]_{AB} \hat{\bs{d}}_{L}^B +
  \hat{\bar{\bs{d}}}_{R}^A \gamma^\mu \left[\hat{g}^d_R\right]_{AB} \hat{\bs{d}}_{R}^B \right) \;,
\end{equation}
with
\begin{equation}
\hat{g}^d_{L,R} = (U^d_{L,R})^\dagger g_{L,R} U^d_{L,R}\;.
\end{equation}

The $Z'$ mediated flavor changing neutral currents (FCNC) between the SM $2\leftrightarrow 3$ generations are naturally suppressed because they only
arise from the mixing with the heavy VL states.

The $Z'$ couplings to neutrinos in the mass basis can be written as
\begin{equation}
  \mathcal L \supset g' Z'_\mu\left(\hat{\bar{\mathbf{N}}}_{L}^\alpha \gamma^\mu \left[\hat{g}^{n}\right]_{\alpha\beta} \hat{\mathbf{N}}_{L}^\beta \right) \;,
\end{equation}
with the coupling
\begin{equation}
 \hat{g}^{n}=\left(U^{\nu}\right)^\mathrm{T}\,g^n\,\left(U^{\nu}\right)^*\;,
\end{equation}
and the gauge basis charge matrix
\begin{equation}
 g^n=\mathrm{diag}\left(0,-1,1,0,0,1,-1,0\right)\;.
\end{equation}

\subsubsection*{Higgs-Lepton Couplings}
The couplings between the physical Higgs boson, $h$, and the charged leptons in the mass basis are
\begin{equation}
  \mathcal L \supset -\frac{1}{\sqrt2}\, h\,\hat{\bar{\bs{e}}}_{L}^A \,\hat{Y}_{AB}^\ell \, \hat{\bs{e}}_{R}^B + \mathrm{h.c.} \;,
\end{equation}
where
\begin{equation}\label{eq:HiggsCouplings}
  \hat Y^\ell = \left(U^\ell_L\right)^\dagger \, Y^\ell \, U^\ell_R \;,
\end{equation}
with the gauge basis couplings
\begin{equation}\label{eq:HiggsCouplingsGauge}
  Y^\ell =
  \begin{pmatrix}
   \lambda_{RL}  & 0 &   0 & 0  \\
   0 & \lambda_{LR} &  0 & 0  \\
    0 & 0 &  y_\tau  & 0  \\
		0 & 0 &  0  & y_\mu \\
  \end{pmatrix} \;.
\end{equation}
A very interesting feature of this model is that the masses of the SM families are to a very high accuracy
linear in the Higgs VEV. Thus, the Higgs couplings in the mass basis, $\hat Y^\ell$, are to a high precision
diagonal in the lower $2\times2$ block.
Hence, the Higgs couplings to the SM states are very much SM-like and there are no significant flavor violating Higgs couplings among the SM states.
We give an analytic proof of this feature in \Appref{app:HiggsCouplings}.
Flavor off-diagonal couplings of the VL states (also to the SM states) can be sizable.

\section{OBSERVABLES}

\subsubsection*{Lepton Non-Universality}

Generally, our model gives rise to lepton non-universality in the operators $\mathcal{O}^{(\prime)}_{i=9,10}$ by tree-level $Z'$ exchange.
The corresponding effective contributions to the Wilson coefficients are
\begin{equation}
 C^{(\prime),\mathrm{NP}}_i=-\frac{\sqrt{2}}{4\,G_\mathrm{F}}\frac{1}{V_{tb} V_{ts}^*} \frac{16\pi^2}{e^2}\frac{1}{2\,v_\Phi^2}\,g^{(\prime)}_{\mathrm{eff},i}\;,
\end{equation}
with the couplings
\begin{align}
 g_{\mathrm{eff},9}&=\left[\hat{g}^d_L\right]_{43}\left[\hat{g}^\ell_R + \hat{g}^\ell_L\right]_{44}\;,& g_{\mathrm{eff},10}&=\left[\hat{g}^d_L\right]_{43}\left[\hat{g}^\ell_R - \hat{g}^\ell_L\right]_{44}\;, & \\
 g^{\prime}_{\mathrm{eff},9}&=\left[\hat{g}^d_R\right]_{43}\left[\hat{g}^\ell_R + \hat{g}^\ell_L\right]_{44}\;,& g^{\prime}_{\mathrm{eff},10}&=\left[\hat{g}^d_R\right]_{43}\left[\hat{g}^\ell_R - \hat{g}^\ell_L\right]_{44}\;.&
\end{align}
These couplings are expressible solely through mixing matrix-elements, for example
\begin{align}
 g_{\mathrm{eff},9}=&
 \left([U^{d\,\dagger}_L]_{43}[U^{d}_L]_{33}-[U^{d\,\dagger}_L]_{42}[U^{d}_L]_{23}\right)\times \\
 &\left([U^{\ell\,\dagger}_L]_{43}[U^{\ell}_L]_{34}-[U^{\ell\,\dagger}_L]_{42}[U^{\ell}_L]_{24}+
 [U^{\ell\,\dagger}_R]_{43}[U^{\ell}_R]_{34}-[U^{\ell\,\dagger}_R]_{42}[U^{\ell}_R]_{24}\right)\;.
\end{align}
While we focus on the $Z'$ coupling to muons in order to explain the observed anomalies, our model also modifies the
effective Wilson coefficients $C^{(\prime),\tau\tau}_{9,10}$ and $C^{(\prime),\nu\nu}_{9,10}$ leading to lepton non-universality
also in $\mathrm{BR}(B_s\to K^{(*)}\tau\bar{\tau})$ and  $\mathrm{BR}(B_s\to \phi\tau\bar{\tau})$.
Quantitative results for these observables have been obtained using the formulas given in \cite{Capdevila:2017iqn} from
where we also adopt the SM prediction (cf.\ also \cite{Bobeth:2011st,Kamenik:2017ghi}).
The NP contributions to the Wilson coefficients $C^{(\prime),\nu\nu}_{9,10}$ affect the SM prediction for $\mathrm{BR}(B_s\to K^{(*)}\nu\bar{\nu})$
and we have followed \cite{Buras:2014fpa, Calibbi:2015kma} to quantify these effects in our model.

\subsubsection*{Muon Anomalous Magnetic Moment}
The $W$, $Z$ and $h$ contributions to the anomalous magnetic moment of the muon are very close to their SM values and we do not detail them here.
On the other hand, the $Z'$ contribution can be sizeable, despite its $\mathrm{TeV}$-scale mass.
This is due to off-diagonal muon-$Z'$ couplings to VL leptons, allowing them to significantly
contribute to the $g-2$ loop.
Since the VL leptons and $Z'$ masses are of the same scale, the leading order contribution has one power of $m_\mu/M_{Z'}$ less than
the naive flavor diagonal $Z'$ contribution (cf.\ e.g.\ the discussion in \cite[sec.\ 7.2]{Jegerlehner:2009ry}).
Parametrically, the dominant modification of $(g-2)_\mu$ in our model is of the size
\begin{equation}
 \delta a^{Z'}_\mu\simeq\frac{m_\mu}{16\pi^2\,v_\Phi}\sum_{a\in\mathrm{VL}}[\hat{g}^\ell_L]_{4a}[\hat{g}^\ell_R]_{4a}\;,
\end{equation}
where the sum goes over the VL leptons.
Naively this points to a scale $v_\Phi\sim10^2\,\mathrm{TeV}$, but the FC couplings to the VL leptons can easily be $\mathcal{O}(0.1)$.
In addition, the contributions of individual VL leptons can partly cancel against one another and we will see
this effect to be at work in our numerical analysis below. We give detailed formulas for $\delta a^{Z'}_\mu$ in \Appref{app:aMuZp}.

\subsubsection*{$\boldsymbol{B_s-\bar{B}_s}$ Mixing}

There is a new tree-level contribution to $B_s-\bar{B}_s$ mixing due to $Z'$ exchange.
We adopt the results and numerical factors from \cite{Altmannshofer:2014cfa, Buras:2012jb} and estimate the relative change of the mixing matrix element
\begin{equation}
 \delta M_{12}\simeq\left(\frac{g^2}{16\pi^2 M_W^2}\left(V_{ts}V_{tb}\right)^2 2.3\right)^{-1}\frac{1}{2\,v_\Phi^2}
 \left(|[\hat{g}^d_L]_{34}|^2+|[\hat{g}^d_R]_{34}|^2+9.7\,\mathrm{Re}([\hat{g}^d_L]_{34}[\hat{g}^{d,*}_R]_{34})\right)\;.
\end{equation}
The most recently updated theoretical uncertainty shows that a deviation from the SM of $\delta M_{12}\lesssim6\%$ can currently not be excluded \cite{DiLuzio:2017fdq}.
This gives an important constraint on the down sector flavor changing $Z'$ couplings.
In our model the $b-s$ coupling is suppressed
because it only arises from the mixing with the heavy VL states such that 
the $Z'$ can be kept at the $\mathrm{TeV}$ scale consistent with the bound derived in \cite{Altmannshofer:2014rta}.

\subsubsection*{Lepton Flavor Violating $\boldsymbol{\tau}$ Decays}
Tree-level $Z'$ exchange also induces the decay $\tau\rightarrow3\mu$. We follow Ref.\ \cite{Okada:1999zk, Kuno:1999jp}
to estimate
\begin{equation}
\begin{split}
 \mathrm{BR(\tau\rightarrow3\mu)}\approx&\frac{1}{\Gamma_\tau}\frac{m_\tau^5}{1536\,\pi^3}\frac{1}{4\,v_\Phi^4} \times \\
 &\left(2\left|[\hat{g}^\ell_L]_{43}[\hat{g}^\ell_L]_{44}\right|^2 + 2\left|[\hat{g}^\ell_R]_{43}[\hat{g}^\ell_R]_{44}\right|^2 +
 \left|[\hat{g}^\ell_L]_{43}[\hat{g}^\ell_R]_{44}\right|^2 + \left|[\hat{g}^\ell_R]_{43}[\hat{g}^\ell_L]_{44}\right|^2\right)\;.
\end{split}
\end{equation}
There is also a new contribution to $\tau\rightarrow\mu\gamma$ due to $Z'$ and the new VL leptons in the loop, which is enhanced by a power of the VL mass.
Using the results of  \cite{Lavoura:2003xp} (cf.\ also \cite{Hisano:1995cp, Ishiwata:2013gma, Abada:2014kba}) we estimate the leading order contribution to be
\begin{equation}
 \mathrm{BR(\tau\rightarrow\mu\gamma)}\simeq\frac{1}{\Gamma_\tau}\frac{\alpha\,m_\tau^3}{1024\,\pi^4}\frac{1}{4\,v_\Phi^2}
 \left\{\left|\sum_{a\in\mathrm{VL}}[\hat{g}^\ell_L]_{4a}[\hat{g}^\ell_R]_{a3}\right|^2+\left|\sum_{a\in\mathrm{VL}}[\hat{g}^\ell_R]_{4a}[\hat{g}^\ell_L]_{a3}\right|^2\right\}\;,
\end{equation}
where the sum is over internal VL leptons. For the numerical analysis we have used a more detailed result which we present in \Appref{app:taumugamma}.

\subsubsection*{Other Observables}
Flavor violating couplings of the $Z$ or the SM scalar $h$ to the SM families are generally suppressed far
below their experimental thresholds. In contrast, flavor changing couplings to the heavy VL
leptons can be large.
Since our $Z'$ is heavy, neutrino trident production \cite{Altmannshofer:2014cfa,Altmannshofer:2014pba}
does not give any important constraints. Lepton unitarity bounds (cf.\ e.g.\ \cite{Fernandez-Martinez:2016lgt, Antusch:2014woa}) are easily fulfilled.
In addition, there are no constraints coming from the branching ratios for $h \rightarrow \gamma \gamma$ or $h \rightarrow g g$ via loop diagrams, since these contributions are suppressed
by factors of  $(v/M_{VL})^2$ (see e.g.\ \cite{delAguila:2008pw, Joglekar:2012vc,Kearney:2012zi,Ishiwata:2013gma}).


\section{ANALYSIS}

\subsubsection*{Strategy}

This model can explain the anomalies in the muon $g-2$ and $b\rightarrow s\mu^+\mu^-$ transitions
without obviously conflicting with other experimental data. To demonstrate this, we have constructed a $\chi^2$-function including the appropriate errors
to simultaneously fit the anomaly in $(g-2)_\mu$ with the value given in \eqref{eq:anomalousg}, and to reproduce
two of the best-fit values of \cite{Altmannshofer:2017fio} 
(cf.\ also \cite{Altmannshofer:2017yso, Alok:2017sui, Capdevila:2017bsm, Ciuchini:2017mik, DAmico:2017mtc, Geng:2017svp, Ghosh:2017ber, Hiller:2017bzc})
for a consistent explanation of $b\rightarrow s\mu^+\mu^-$ anomalies:
\begin{align}\label{eq:GlobalNPfits}
\mathrm{I.}&\,:& C^{\mathrm{NP}}_9&\approx-1.21\pm0.2\;,& C'_9&\approx C^{\mathrm{NP}}_{10}\approx C'_{10}\approx0\;,& \\
\mathrm{or} \quad \mathrm{II.}&\,:& C^{\mathrm{NP}}_9&\approx-1.25\pm0.2\;,& C'_9&\approx0.59\pm0.2\;,&  C^{\mathrm{NP}}_{10}&\approx C'_{10}\approx0\;.&
\end{align}
Furthermore we fit the masses $m_d$, $m_s$, $m_\mu$, and $m_\tau$ at the weak scale \cite{Antusch:2013jca} while requiring $\delta M_{12}\lesssim15\%$
to be consistent with the data and theoretical error of $B_s-\bar{B}_s$ mixing.
All other observables are not constrained in the fit, i.e.\ they arise as predictions of our best fit points.
However, there are currently more parameters than observables and so it is not excluded that there are more points that fit the data well
with different predictions. The first family could always be included in the analysis in a straightforward way without affecting our conclusions.\footnote{%
Including the first family in the most straightforward way the Yukawa couplings and hence resulting mass matrices are extended to (and similarly for neutrinos and down quarks) 
\begin{align}
  \mathcal{M}^\ell &=
 \begin{pmatrix}
    \lambda_{RL}\,v & \lambda_E\,v_\Phi &   \lambda_3\,v_\Phi & \lambda_2\,v_\varphi & 0 \\
    \lambda_L\,v_\Phi & \lambda_{LR}\,v &  0 & 0 & 0 \\
    \lambda_3 \,v_\Phi &  0 &  y_\tau\,v & 0 & 0 \\
   \lambda_2\,v_\varphi &  0 &  0 &  y_{22}\,v  & y_{21}\,v \\
   0 &  0 &  0 &  y_{12}\,v & y_{11}\,v \\
  \end{pmatrix}\;.
\end{align}
}

\subsubsection*{Results}
We have found two points which give a very good fit for the cases (I.) and (II.), they are listed in \Tabref{tab:bestfit} (and in \Tabref{tab:bestfitII} in \Appref{tab:bestfitII}).
We cannot find a good fit for $C^\mathrm{NP}_9\simeq-C^\mathrm{NP}_{10}$.

\begin{table}[t]
  \begin{center}\setlength{\tabcolsep}{2pt}
    \begin{tabular}{lll@{\hskip 12pt}lll}
      \toprule
   \multicolumn{6}{c}{Best fit point I.} \\\midrule
$\lambda_\mu$ &=& $-0.00008520346$ & 
$\lambda_\tau$ &=& $\phantom{-}0.010031941$ \\ 
$\lambda_s$ &=& $-0.002313419$ & 
$\lambda_b$ &=& $-0.01639676$ \\ 
$\lambda_L$ &=& $\phantom{-}0.8595483$ & 
$\lambda_E$ &=& $\phantom{-}0.8596570$ \\ 
$\lambda_D$ &=& $-1.7819031$ & 
$\lambda_Q$ &=& $-0.3253384$ \\ 
$\lambda_3$ &=& $-0.018705093$ & 
$\lambda_2$ &=& $-0.7701059$ \\ 
$\lambda_{RL}$ &=& $-0.9926161$ & 
$\lambda_{LR}$ &=& $\phantom{-}0.0014518601$ \\
$v_\varphi$ &=& $\phantom{-}1603.0788$ & 
$v_\phi$ &=& $\phantom{-}1622.5729$ \\ 
$g'$ &=& $-0.7518533$ \\\bottomrule 
    \end{tabular}
  \end{center}
  \caption{Best fit point to the data.
  We do not list $M_R$, $\lambda_{\nu 1,2}$ and $\lambda_N$ because their precise values do not affect the results.}
  \label{tab:bestfit}
\end{table}

The predictions of the best fit points are listed in \Tabref{tab:results} together with current experimental bounds.
Effects on other observables such as $h\to\mu\tau$, $h\to\gamma\gamma$,
$h\to gg$, neutrino trident production or PMNS unitarity violation have also been considered, but they are robustly suppressed
in this model and so we do not discuss them in detail. While $Z\to\mu\tau$ is practically absent at tree level,
the dominant contribution arises from a one-loop diagram involving
$Z'$ and VL leptons in the loop. A rough estimate of this shows that $\mathrm{BR}(Z\to\mu\tau)$
nonetheless comes out well below the current bound of $1.2\times10^{-5}$ \cite{Olive:2016xmw}.

\begin{table}[t]
\centering
\begin{tabular}{cccc}
      \toprule
      Observable & Best Fit I. & Best Fit II. & Bound \\
      \midrule
     $m_L,m_E$ & $1.78\,\mathrm{TeV}$, $1.95\,\mathrm{TeV}$ & $2.08\,\mathrm{TeV}$, $2.25\,\mathrm{TeV}$  & $> 450~\mathrm{GeV}$ \cite{Falkowski:2013jya}\\
     $m_Q,m_D$ & $1.34\,\mathrm{TeV}$, $3.15\,\mathrm{TeV}$ & $1.61\,\mathrm{TeV}$, $3.49\,\mathrm{TeV}$  & $> 900~\mathrm{GeV}$ \cite{Khachatryan:2015gza}\\
     $M_D$     & $1.86\,\mathrm{TeV}$                       & $2.16\,\mathrm{TeV}$                        & \\
     $M_{Z'}$  & $1.73\,\mathrm{TeV}$                       & $2.25\,\mathrm{TeV}$                        & \\
     $\tau\to\mu\gamma$ & $4.6\times10^{-9}$    & $5.3\times10^{-9}$ & $<4.4\times10^{-8}$ \cite{Aubert:2009ag} \\
     $\tau\to3\mu$      & \;$6.7\times10^{-10}$ & $1.1\times10^{-9}$ & $<2.1\times10^{-8}$ \cite{Hayasaka:2010np} \\
     $\delta M_{12} (B_s-\bar{B}_s)$ & $1\%$ & $-3\%$     & $\lesssim\pm6\%$ \cite{DiLuzio:2017fdq} \\
     $\mathrm{BR}(B_s\to K\tau\bar{\tau})^{[15,22]}$         & $1.8\times10^{-7}$ & $1.4\times10^{-7}$ & \;$<2.25\times10^{-3}$ \cite{TheBaBar:2016xwe} \\
     $\mathrm{BR}(B_s\to K^{*}\tau\bar{\tau})^{[15,19]}$\!\! & $2.5\times10^{-7}$ & $2.9\times10^{-7}$ &  \\
     $\mathrm{BR}(B_s\to \phi\tau\bar{\tau})^{[15,18.8]}$\!  & $2.2\times10^{-7}$ & $2.6\times10^{-7}$ &  \\
     $R_K^{\nu\bar{\nu}}$ & $0.91$ & $0.93$ & $<4.3$ \cite{Lees:2013kla} \\
     $R_{K^*}^{\nu\bar{\nu}}$ & $0.91$ & $0.93$ & $<4.4$ \cite{Lutz:2013ftz} \\
      \bottomrule
    \end{tabular}
\caption{Values of (unfitted) observables at our best fit points (central values) and corresponding bounds.}\label{tab:results}
\end{table}

\begin{figure}[!ht]
\captionsetup[subfigure]{justification=centering}
\centering
\begin{subfigure}[]{.45\textwidth}
\centering
\includegraphics[width=1\textwidth]{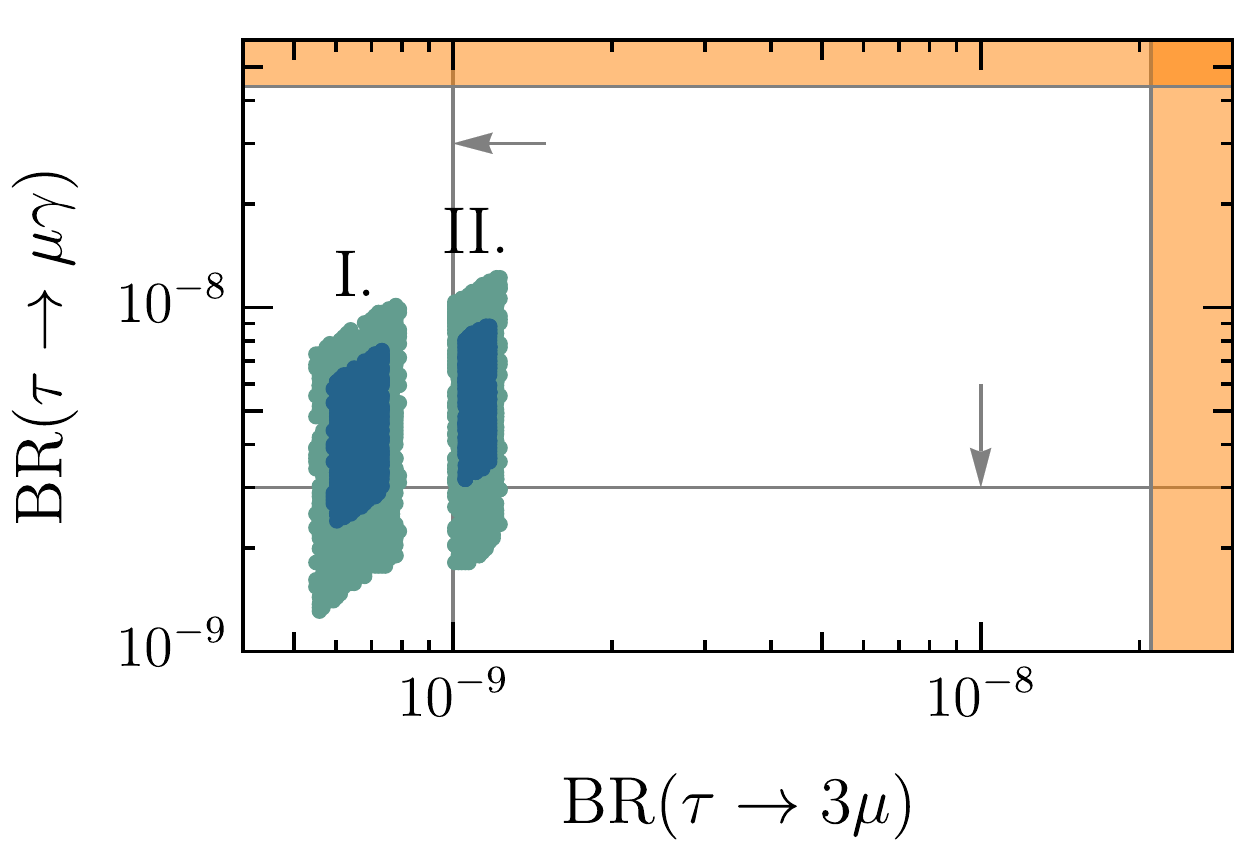}
\caption{}
\label{fig:a}
\end{subfigure}
\hspace{1cm}
\begin{subfigure}[]{.45\textwidth}
\centering
\includegraphics[width=1\textwidth]{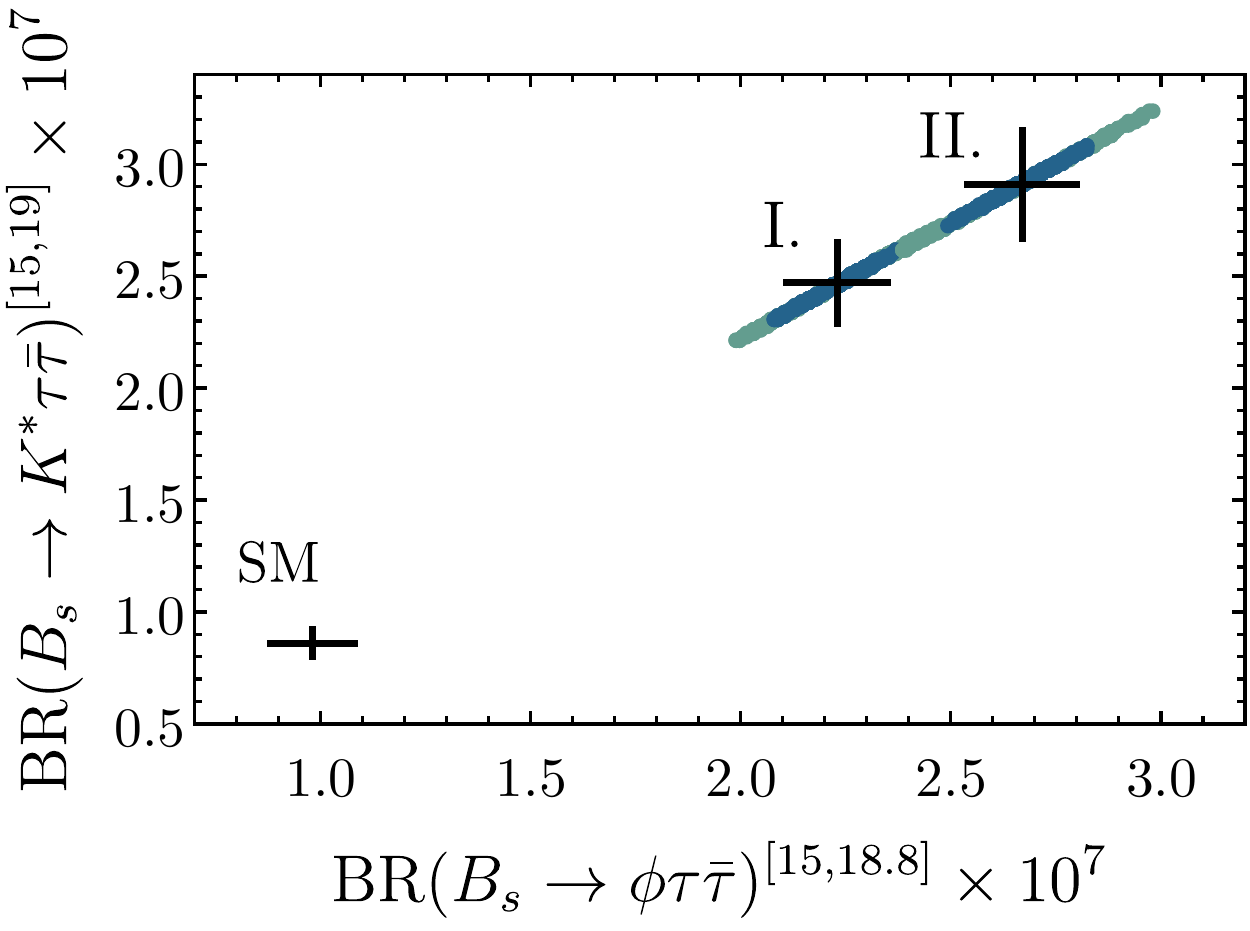}
\caption{}
\label{fig:b}
\end{subfigure}%
\vspace{0.5cm}
\begin{subfigure}{.45\textwidth}
\centering
\includegraphics[width=1\textwidth]{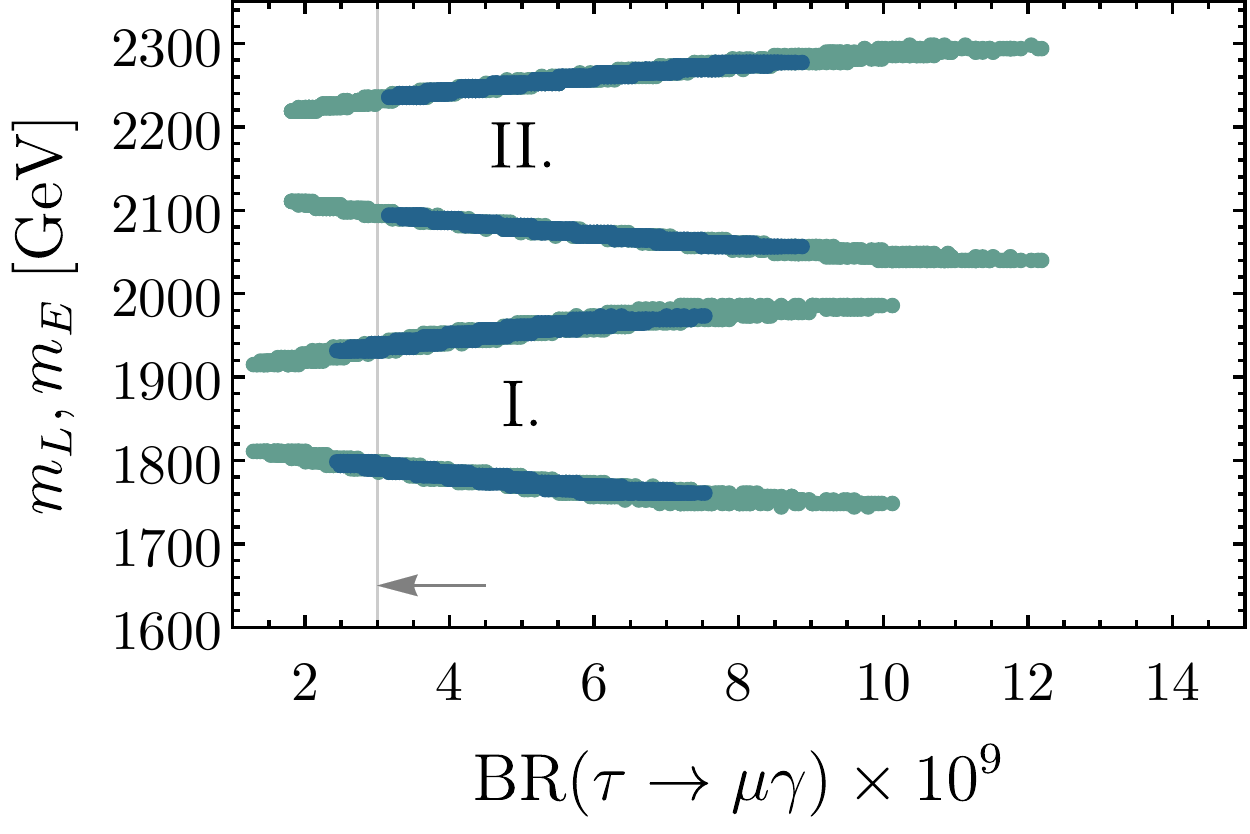}
\caption{}
\label{fig:c}
\end{subfigure}
\hspace{1cm}
\begin{subfigure}{.45\textwidth}
\centering
\includegraphics[width=1\textwidth]{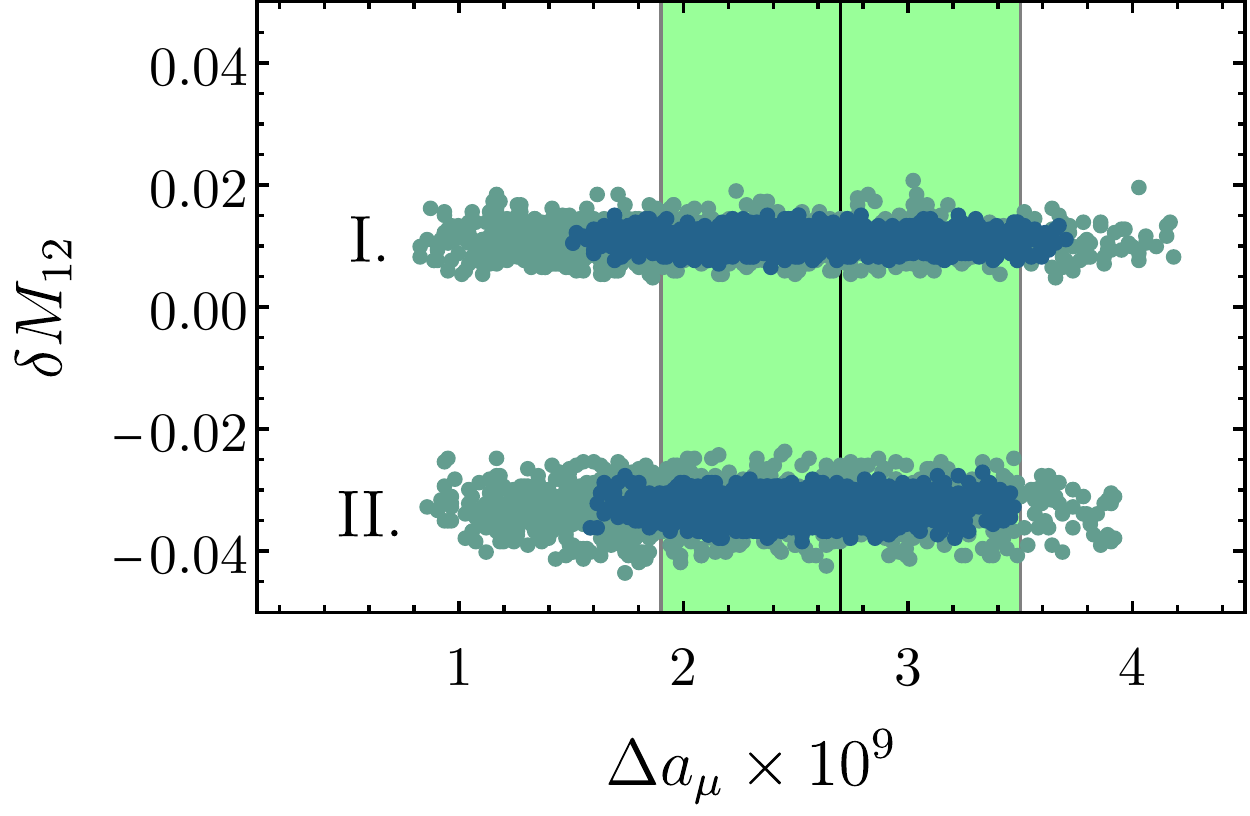}
\caption{}
\label{fig:d}
\end{subfigure}
\caption{The dots show random events scattered in the $\chi^2<1(3)$ regions around our best fit points (I.) (\Tabref{tab:bestfit}) and (II.) (\Tabref{tab:bestfitII}).
The colored regions and gray lines in \Figref{fig:a} and \Figref{fig:c} show the current experimental exclusion \cite{Aubert:2009ag,Hayasaka:2010np} and prospects \cite{Aushev:2010bq}, respectively.
The lower cross in \Figref{fig:b} denotes the SM expectation \cite{Capdevila:2017iqn} while the upper crosses show our best fit points with errors computed as in \cite{Capdevila:2017iqn}.
The green area in \Figref{fig:d} shows the best fit value and errors as in eq.\ \eqref{eq:anomalousg}.}
\label{fig:Plots}
\end{figure}

The loop contributions to $(g - 2)_\mu$ from $Z$, $W$, and $h$ are very close to their SM values while the FC $Z'$ exchange with VL leptons in the loop completely accounts for the anomaly.
The $Z'$ contribution to $B_s-\bar B_s$ is mixing suppressed while larger flavor diagonal couplings to $\mu\bar{\mu}$ can explain a sizable $C^\mathrm{NP}_9$.
The fit prefers a corner of the parameter space where $\lambda_E\sim\lambda_L$ which makes the
$Z'$ couplings to leptons approximately left-right (anti-)symmetric (the couplings are LR symmetric or LR anti-symmetric depending on the specific coupling and 
we display the coupling matrices for one of the best fit points in \Appref{app:bestfit}).
The approximately LR symmetric couplings to the mu and tau leptons leads to an enhanced contribution to $C^\mathrm{NP}_9$ (and possibly also $C^\prime_9$)
and a dramatic cancellation in the axial vector couplings to leptons $C^{(\prime)}_{10}$.
By contrast, the approximately equal but opposite LR \textit{(anti-)}symmetric contributions of individual 
VL leptons to the muon $g-2$ and $\tau\to\mu\gamma$ cancel only to an order of magnitude.
There are no cancellations in $\tau\to3\mu$ where the couplings, in fact, add up constructively.
Nevertheless, the tree level process $\tau\to3\mu$ is suppressed by the naturally small flavor off-diagonal
couplings of the $Z'$ to the SM fermion generations. The best-fit predictions for $\mathrm{BR}(\tau\to\mu\gamma)$ and $\mathrm{BR}(\tau\to3\mu)$ fall
close to regions that can be probed by future experiments \cite{Aushev:2010bq}, cf.\ \Figref{fig:a}.
A positive value of $C^{\prime}_9$ allows for a negative shift of $\delta M_{12}$, thereby cushioning a
$1.8\sigma$ tension between SM and experiment \cite{DiLuzio:2017fdq}. \enlargethispage{14pt}
This happens to be the case for our best fit point (II.).

The NP contributions to the Wilson coefficients $C^{(\prime),\tau\tau}_{9,10}$ qualitatively follows the patterns in \eqref{eq:GlobalNPfits} but with reversed signs.
$\mathrm{BR}(B_s\to K\tau\bar{\tau})$ and  $\mathrm{BR}(B_s\to \phi\tau\bar{\tau})$ are significantly increased compared to the SM, cf.\ \Figref{fig:b}.
By contrast $R_{K^{(*)}}^{\nu\bar{\nu}}$ is suppressed compared to the SM.
The suppression of $R_{K^{(*)}}^{\nu\bar{\nu}}$ together with a $Z'$ explanation of $b\rightarrow s\mu^+\mu^-$ is not in contradiction with the results of \cite{Buras:2014fpa}.
In our model, we find that the enhancement of $R_{K^{(*)}}^{\nu\bar{\nu}}$ from the second family \cite{Buras:2014fpa} is counter-acted by a vast suppression of the third family contributions.

We do not find any observable that causes problems. We stress again that the $Z'$
does, by construction, only very feebly couple to the SM first generation.
The gauge coupling $g'$ is only very mildly constrained by the fit with a
$\chi^2<3$ region of $g'\in-\left[5.33(5.08),0.32(0.39)\right]$ for point I.(II.).
Bounds for family specific $Z'$ bosons can be obtained from the LHC $\sqrt{s}=13\,\mathrm{TeV}$ searches \cite{Khachatryan:2016zqb,Aaboud:2017buh}
and range from $M_{Z'}\gtrsim1.3-2.0\,\mathrm{TeV}$ \cite{Alonso:2017uky,Bonilla:2017lsq}, while some models can survive with $M_{Z'}$ as low as
$\gtrsim500\,\mathrm{GeV}$ \cite{Bian:2017rpg}. Deriving a robust limit for our model requires to specify the details of the first family couplings which is beyond the scope of this 
work. Nevertheless, our $Z'$ mass comes out right in the ballpark of current limits and so it can be searched for at the LHC in dimuon final states. 
If not with the full run 2 data, our best fit parameter regions will conclusively be tested by a high-luminosity run of the LHC.

We have scattered one thousand points randomly within the $\chi^2<1(3)$ regions around the best fit values.
There are strong correlations among the masses of the pair of VL leptons, and between the VL lepton masses
and the prediction for $\mathrm{BR}(\tau\to\mu\gamma)$ \Figref{fig:c}.
Also $\mathrm{BR}(B_s\to K\tau\bar{\tau})$ and  $\mathrm{BR}(B_s\to \phi\tau\bar{\tau})$
are strongly correlated \Figref{fig:b}. All other predicted observables are largely uncorrelated, see e.g.\ \Figref{fig:d}.

Given the parameters of our best fit points we can also evaluate the effective Higgs couplings of
the SM quarks and leptons (see \Appref{app:HiggsCouplings} for details).
The resulting Higgs couplings are diagonal in the mass basis and,
moreover, proportional to the masses, just like in the Standard Model.

For model building it is interesting to compare the (gauge basis) Higgs Yukawa couplings
to the mixing induced contributions to masses of the SM fermion generations.
The muon and strange quark obtain mass predominantly due to mixing effects,
and this is also reflected by their large off--diagonal couplings to the VL states.
The tau lepton and b quark, by contrast, obtain almost all of their mass from the direct Higgs Yukawa
coupling and mixing effects are miniscule.


\section{CONCLUSION}

We have studied the Standard Model extended by one complete family of ``VL'' fermions, including right-handed neutrinos,
which are vector-like with respect to the Standard Model gauge group.
In addition we have introduced a new ``$\U1_{3-4}$'' gauge symmetry under which the third SM family of quarks and leptons have
$\U1_{3-4}$ charge $1$ while the left-chiral part of the new 4th family has charge $-1$.
Hence, the model is free of gauge anomalies.
In our analysis, we have only considered the mixing of the VL states with the second and third SM families,
while the absence of mixing to the first family is motivated by a high scale flavor symmetry.
The first family could always be included in a straightforward way without affecting our conclusions, and this should be done in the future
in order to study the collider phenomenology of this model in more detail.

We have shown that this model can fit the anomalies in the muon $g-2$ and $b\rightarrow s\mu^+\mu^-$ transitions
without conflicting with other experimental data.
The best fit points predict (cf.\ \Tabref{tab:results}) new quarks, leptons and a family specific $Z'$ at the $\mathrm{TeV}$ scale, as well as testable effects
in the lepton flavor violating processes $\tau\to\mu\gamma$ and $\tau\rightarrow3\mu$.
Furthermore, we find a significant enhancement of $\mathrm{BR}(B_s\to K^{(*)}\tau\bar{\tau})$ and $\mathrm{BR}(B_s\to \phi\tau\bar{\tau})$, while $R_{K^{(*)}}^{\nu\bar{\nu}}$
is suppressed relative to the SM.
Effects on other observables such as $Z\to\mu\tau$, $h\to\mu\tau$, $h\to\gamma\gamma$, $h\to gg$, neutrino trident production,
or PMNS unitarity violation are all robustly suppressed.
The Higgs couplings in our model are to a very high degree SM-like and we have given an analytic proof of that.


\appendix
\section{Neutrino See-Saw}\label{app:NeutrinoMasses}
The neutrino mass matrix is described in \eqref{eq:nuMasses}.
In the limit $M\sim M_{L,R}\gg v_{\Phi,\varphi}\gg v$ the eigenvectors of $\mathcal{M}^\nu$ are to a very good approximation given by
\begin{align}
 N_{1,2}=&\left(0,0,0,0,0,\frac{1}{\sqrt{2}},\pm\frac{1}{\sqrt{2}},0\right)\;, \\
 N_{3,4}=&\left(\frac{1}{\sqrt{2}},0,0,0,0,0,0,\pm\frac{1}{\sqrt{2}}\right)\;, \\
 N_{5,6}=&\frac{1}{\sqrt{2}}\left(0,\frac{\lambda_L\,v_\Phi}{M_D},\frac{\lambda_3\,v_\Phi}{M_D},\frac{\lambda_2\,v_\varphi}{M_D},\pm1,0,0,0\right)\;, \\
 N_{7}=&\left(0,\frac{\lambda_3}{\sqrt{\lambda_3^2+\lambda_L^2}},-\frac{\lambda_L}{\sqrt{\lambda_3^2+\lambda_L^2}},0,0,0,0,0\right)\;, \\
 N_{8}=&\frac{1}{\mathcal{N}_8}\left(0,1,-\frac{\lambda_3}{\lambda_L},-\frac{v_\Phi}{v_\varphi}\frac{\left(\lambda_3^2+\lambda_L^2\right)}{\lambda_2\,\lambda_L},0,0,0,0\right)\;.
 \end{align}
Here we have used
\begin{align}
M_D:=&\sqrt{\left(\lambda_L\,v_\Phi\right)^2 + \left(\lambda_3\,v_\Phi\right)^2 + \left(\lambda_2\,v_\varphi\right)^2}\;, \\
\mathcal{N}_8:=&\sqrt{1+\left(\frac{\lambda_3}{\lambda_L}\right)^2+\left[\frac{v_\Phi}{v_\varphi}\frac{\left(\lambda_3^2+\lambda_L^2\right)}{\lambda_2\,\lambda_L}\right]^2}\;.
\end{align}
It is then straightforward to construct the diagonalization matrix $U^\nu$ from the eigenvectors.

\section{Higgs Coupling Diagonalization}\label{app:HiggsCouplings}
We wish to show that the Higgs couplings \eqref{eq:HiggsCouplings} are diagonal in the mass basis.
We will focus on the charged leptons here, but the whole analysis of this section fully applies also to the d-type quarks by formally replacing
$\lambda_E\rightarrow\lambda_D$, $\lambda_L\rightarrow\lambda_Q$, $y_\tau\rightarrow y_b$, and $y_\mu\rightarrow y_s$.
Phenomenologically we are led to the relations $v\ll v_\Phi\sim v_\varphi$ and $\lambda_{LR},y_\tau,y_\mu\ll1$.
Therefore, we will treat the lower $3\times3$ block of $\mathcal{M}^\ell$ in \eqref{eq:leptonmass} as perturbation.
The leading order mass matrix, hence, is given by
\begin{equation}
\mathcal{M}^\ell\approx \widetilde{\mathcal{M}}^\ell\equiv v_\Phi
\begin{pmatrix}
    \lambda_{RL}\,\xi & \lambda_E &   \lambda_3 & \lambda_2 \\
    \lambda_L & 0 &  0 & 0 \\
    \lambda_3 &  0 &  0 & 0 \\
   \lambda_2 &  0 &  0 &  0 \\
  \end{pmatrix}\;,
\end{equation}
where $\xi:=v/v_\Phi$. The left- and right-singular vectors of this matrix are given by
\begin{align}
u^\ell_{L,1}=&\frac{1}{N^{1/2}_{L,1}}\left(-u^{L}_{-},\frac{\lambda_L}{\lambda_2},\frac{\lambda_3}{\lambda_2},1\right)\;,& u^\ell_{R,1}=&\frac{1}{N^{1/2}_{R,1}}\left(u^{R}_{-},-\frac{\lambda_E}{\lambda_2},-\frac{\lambda_3}{\lambda_2},-1\right)\;,& \\
u^\ell_{L,2}=&\frac{1}{N^{1/2}_{L,2}}\left(u^{L}_{+},-\frac{\lambda_L}{\lambda_2},-\frac{\lambda_3}{\lambda_2},-1\right)\;,& u^\ell_{R,2}=&\frac{1}{N^{1/2}_{R,2}}\left(u^{R}_{+},-\frac{\lambda_E}{\lambda_2},-\frac{\lambda_3}{\lambda_2},-1\right)\;,& \\
u^\ell_{L,3}=&\frac{1}{N^{1/2}_{L,3}}\left(0,-\frac{\lambda_3}{\lambda_L},1,0\right)\;,& u^\ell_{R,3}=&\frac{1}{N^{1/2}_{R,3}}\left(0,-\frac{\lambda_3}{\lambda_E},1,0\right)\;,& \\
u^\ell_{L,4}=&\frac{1}{N^{1/2}_{L,4}}\left(0,-\frac{\lambda_2}{\lambda_L},0,1\right)\;,& u^\ell_{R,4}=&\frac{1}{N^{1/2}_{R,4}}\left(0,-\frac{\lambda_2}{\lambda_E},0,1\right)\;,&
\end{align}
where
\begin{equation}
\scalebox{1.15}{
$u^{L(R)}_{\mp}:=\frac{(-)\left(\lambda_L^2-\lambda_E^2\right)-\xi^2\lambda_{RL}^2\mp\sqrt{
\left(\lambda_E^2-\lambda_L^2\right)^2+4 \xi^2 \lambda_2^2 \lambda_{RL}^2 + 4 \xi^2 \lambda_3^2 \lambda_{RL}^2 + 2\xi^2\lambda_E\lambda_{RL}^2+2 \xi^2\lambda_L^2 \lambda_{RL}^2+\xi^4 \lambda_{RL}^4}}
{2\,\xi\,\lambda_2\,\lambda_{RL}}\;,$}
\end{equation}
\normalsize
and the normalization factors $N_{L(R),i}$ are defined by the requirement $|u^\ell_{L(R),i}|^2=1$.
The matrices $\widetilde{U}^\ell_{L,R}$ have these vectors as columns.
The four corresponding singular values are given by
\begin{equation}
  (\widetilde{U}^\ell_L)^\dagger \, \widetilde{\mathcal{M}}^\ell \, \widetilde{U}^\ell_R=\mathrm{diag}\left( \widetilde{m}_{E}, \widetilde{m}_{L},0, 0\right)\;,
\end{equation}
where one can obtain analytic expressions for $\widetilde{m}_{E,L}$ but we do not need them here and so they are not displayed.
Numerically, $\widetilde{U}^\ell_{L,R}\approx U^\ell_{L,R}$ and $\widetilde{m}_{E}\approx{m}_{E}$ and $\widetilde{m}_{L}\approx{m}_{L}$ hold at the percent level.
Now regard the perturbation
\begin{equation}
\mathcal{M}^\ell=\widetilde{\mathcal{M}}^\ell+\xi\,\mathrm{diag}\left( 0, \lambda_{LR}, y_\tau, y_\mu \right)\;.
\end{equation}
To leading order in perturbation theory the eigenvalues of the perturbed matrix are obtained by diagonalizing it with the
zeroth-order left- and right-singular matrices.
Following this procedure we find the small singular values of $\mathcal{M}^\ell$ to be
\begin{align}
\widetilde{m}_\tau&=\frac{v}{N_3}\left|\lambda_\tau+\frac{\lambda_3^2\,\lambda_{LR}}{\lambda_E\,\lambda_L}\right|\,,& &\mathrm{and}&
\widetilde{m}_\mu&=\frac{v}{N_2}\left|\lambda_\mu+\frac{\lambda_2^2\,\lambda_{LR}}{\lambda_E\,\lambda_L}\right|\,,&
\end{align}
with
\begin{equation}
N_{2(3)}:=\sqrt{\left[1+\left(\lambda_{2(3)}/\lambda_E\right)^2\right]\left[1+\left(\lambda_{2(3)}/\lambda_L\right)^2\right]}\;.
\end{equation}
These values agree with the numerical values to $\mathcal{O}(1\%)$. There are off-diagonal residuals which are given by
\begin{align}
\left[(\widetilde{U}^\ell_L)^\dagger \, \mathcal{M}^\ell \, \widetilde{U}^\ell_R\right]_{34}~=&~\delta m_{\tau\mu}~\equiv~
\frac{v\,\lambda_2\,\lambda_3\,\lambda_{LR}}{\sqrt{\left(\lambda_2^2+\lambda^2_E\right)\left(\lambda_{3}^2+\lambda^2_L\right)}}\;, \\
\left[(\widetilde{U}^\ell_L)^\dagger \, \mathcal{M}^\ell \, \widetilde{U}^\ell_R\right]_{43}~=&~\delta m_{\mu\tau}~\equiv~
\frac{v\,\lambda_2\,\lambda_3\,\lambda_{LR}}{\sqrt{\left(\lambda_3^2+\lambda^2_E\right)\left(\lambda_{2}^2+\lambda^2_L\right)}}\;.
\end{align}
They are numerically subdominant compared to the diagonal entries.

The crucial point is that the Higgs couplings in the gauge basis \eqref{eq:HiggsCouplingsGauge}
are, at least with regard to the lower $2\times2$ block, exactly of the same form as the perturbations to the mass matrix.
Approximately diagonalizing the mass matrices with $\widetilde{U}^\ell_{L,R}$, hence, results in the Higgs couplings
\begin{equation}\label{eq:HiggsCouplings2x2}
  \widetilde Y^\ell_{\mathrm{light}} = \frac{1}{v}
  \begin{pmatrix}
   \widetilde{m}_\tau  & \delta m_{\tau\mu} \\
   \delta m_{\mu\tau} & \widetilde{m}_\mu
  \end{pmatrix}\;,
\end{equation}
which are directly proportional to the mass matrices.
There are slight deviations to the proportionality arising at higher order, but they are numerically not relevant.

Evaluating the effective Higgs couplings in the mass basis numerically for our best-fit point (I.) (cf.\ \Tabref{tab:bestfit}) one finds
\begin{align}  \label{eq:ULe2}
  \hat{Y}^{\ell}_{\mathrm{light}} &= \frac{\mathrm{GeV}}{v}
  \begin{pmatrix}
1.74618 & 3.46843\times10^{-7}  \\
3.47055\times10^{-7}  & 0.102717  \\
  \end{pmatrix}\;.& \\
\hat{Y}^{d}_{\mathrm{light}} &= \frac{\mathrm{GeV}}{v}
  \begin{pmatrix}
2.85391 & 7.71892\times10^{-7}  \\
1.41639\times10^{-6}  & 0.0543716  \\
  \end{pmatrix}\;.&
  \end{align}
Clearly the off-diagonal corrections are negligible. Thus, the Higgs coupling to quarks and leptons
is diagonal in the mass basis and, moreover, proportional to the masses, just like in the Standard Model.

\subsubsection*{Integrating out the VL Fermions}
The fact that the Higgs couplings are very much SM like also holds for the effective mass matrix of the light (SM) fermions
which is obtained after integrating out the heavy VL fermions, and we wish to show this analytically.
The $4 \times 4$ mass matrix \eqref{eq:leptonmass} can be rotated to a basis where there is a heavy $2 \times 2$ block.
This is given by the transformation
\begin{equation}
 M^\ell = \left(X_L\right)^\dagger \, \mathcal{M}^\ell \, X_R\;,
\end{equation}
with
\begin{equation}
X_L = \begin{pmatrix}
   1 &  0 & \phantom{-}\,0 & 0   \\
   0 & N_{2L} \,\lambda_L v_\Phi & \phantom{-}N_{3L}\,  \lambda_3 v_\Phi & N_{4L} \lambda_L v_\Phi \\
   0 & N_{2L} \,\lambda_3 v_\Phi & -N_{3L}\, \lambda_L v_\Phi & N_{4L} \lambda_3 v_\Phi\\
   0 & N_{2L} \,\lambda_2 v_\varphi & \phantom{-}\,0 &  - N_{4L} (\lambda_L^2 + \lambda_3^2)v_\Phi
  \end{pmatrix}\;,
\end{equation}
and
\begin{equation}
 X_R = \begin{pmatrix}
   1 & 0 & 0 & 0   \\
   0 & N_{2R} \, \lambda_E v_\Phi & \phantom{-}N_{2R} \, \lambda_3 v_\Phi & N_{2R} \, \lambda_2 v_\varphi \\
   0 &  N_{3R} \, \lambda_3 v_\Phi &  - N_{3R} \, \lambda_E v_\Phi & 0 \\
   0  & N_{4R} \, \lambda_E v_\Phi  &   \phantom{-}N_{4R} \, \lambda_3 v_\Phi & -N_{4R} \, (\lambda_E^2 + \lambda_3^2)v_\Phi
  \end{pmatrix}\;,
\end{equation}
where
\begin{align}
 N_{2L(R)} := & \left[ \left(\lambda_{L(E)}^2 +  \lambda_3^2 \right) v_\Phi^2 +  \lambda_2^2\, v_\varphi^2\right]^{-1/2}\;,&  \\
 N_{3L(R)} := & \left[ \left(\lambda_{L(E)}^2 +  \lambda_3^2 \right) v_\Phi^2 \right]^{-1/2}\;,& \\
 N_{4L(R)} := & \left[ \left(\lambda_{L(E)}^2 + \lambda_3^2\right) \lambda_2^2v_\varphi^2 + \left(\lambda_{L(E)}^2 + \lambda_3^2\right)^2v_\Phi^2\right]^{-1/2}\;.&
\end{align}
In terms of $2 \times 2$ blocks $M^\ell$ is then given by
  \begin{equation}
  M^\ell =
  \begin{pmatrix}
  M_{\mathrm{heavy}} & V_R   \\
  V_L & M_{\mathrm{light}}
  \end{pmatrix}\;.
\end{equation}
Upon integrating out the heavy states perturbatively we obtain the effective $ 2\times2$ mass matrix for the light states given by
\begin{equation}
 M^{\mathrm{eff}}_{\mathrm{light}} = M_{\mathrm{light}} - V_L \, M_{\mathrm{heavy}}^{-1} \ V_R\;.
\end{equation}
In particular one finds that the effective light masses are to very high accuracy linear in the Higgs VEV.
There are higher order corrections in $v$ but they are numerically irrelevant.
Thus when one diagonalizes the fermion mass matrix one simultaneously diagonalizes the coupling of the Higgs field to fermions.
Hence there are no significant flavor violating Higgs couplings.

\section{Details of $\boldsymbol{\delta a^{Z'}_\mu}$}\label{app:aMuZp}
The leading order contribution to the muon $g-2$ arising from the $Z'$ coupling to leptons as in \eqref{eq:ZpLagrangian}
is given by (see e.g.\ \cite{Jegerlehner:2009ry,Dermisek:2013gta})
\begin{equation}
 \delta a^{Z'}_\mu=-g^{\prime\,2}\frac{\,m_\mu^2}{8\pi^2\,M_{Z'}^2}
 \sum_a\left[ \left( |[\hat{g}^\ell_L]_{4a}|^2+|[\hat{g}^\ell_R]_{4a}|^2\right)F(x_a) +
              \mathrm{Re}([\hat{g}^\ell_L]_{4a}[{\hat{g}^{\ell,*}}_R]_{4a})\,\frac{m_a}{m_\mu}\,G(x_a)\right]\;,
\end{equation}
where $a$ runs over all leptons with mass $m_a$ in the loop, $x_a:=(m_a/M_{Z'})^2$, and the loop functions
\begin{align}\label{eq:loopF}
 F(x)&:=\left(5\,x^4 - 14\,x^3 + 39\,x^2 - 38\,x - 18\,x^2 \ln x + 8\right)/\left[12 \left(1 - x\right)^4\right]\;, \\\label{eq:loopG}
 G(x)&:=\left(x^3 + 3\,x - 6\,x \ln x - 4\right)/\left[2 \left(1 - x\right)^3\right]\;.
\end{align}

\section{Details of $\boldsymbol{\tau\rightarrow\mu\gamma}$}\label{app:taumugamma}
The leading order contribution to $\tau\rightarrow\mu\gamma$ arises from the flavor off-diagonal $Z'$ couplings
between $\tau-\mathrm{VL}$ and $\mu-\mathrm{VL}$. We have used the general results given in \cite{Lavoura:2003xp} to find the leading order contributions to
the partial width
\begin{equation}
 \Gamma(\tau\rightarrow\mu\gamma)=\frac{\alpha\,g^{\prime\,4}}{1024\,\pi^4}\frac{m_\tau^5}{M_{Z'}^4}\left(\left|\tilde\sigma_L\right|^2+\left|\tilde\sigma_R\right|^2\right)\;,
\end{equation}
with
\begin{align}
 \tilde\sigma_L=&\sum_a \left([\hat{g}^\ell_L]_{4a} [\hat{g}^\ell_L]_{a3}\,F(x_a)+ \frac{m_a}{m_\tau}\,[\hat{g}^\ell_L]_{4a} [\hat{g}^\ell_R]_{a3}\,G(x_a) \right)\;, \\
 \tilde\sigma_R=&\sum_a \left([\hat{g}^\ell_R]_{4a} [\hat{g}^\ell_R]_{a3}\,F(x_a)+ \frac{m_a}{m_\tau}\,[\hat{g}^\ell_R]_{4a} [\hat{g}^\ell_L]_{a3}\,G(x_a) \right)\;,
\end{align}
where $a$ runs over all leptons with mass $m_a$ in the loop, $x_a:=(m_a/M_{Z'})^2$, and the loop functions are the same as in \eqref{eq:loopF} and \eqref{eq:loopG}.

\section{$\boldsymbol{Z'}$ Couplings at the Best Fit Point(s)}\label{app:bestfit}
\begin{table}[t]
  \begin{center}\setlength{\tabcolsep}{2pt}
    \begin{tabular}{lll@{\hskip 12pt}lll}
      \toprule
       \multicolumn{6}{c}{Best fit point II.}  \\\midrule
$\lambda_\mu $ &=& $ -0.00023171652$ & 
$\lambda_\tau $ &=& $ \phantom{-}0.010033469$ \\ 
$\lambda_s $ &=& $ -0.0009612744$ & 
$\lambda_b $ &=& $ -0.016371133$ \\ 
$\lambda_L $ &=& $ \phantom{-}0.8806972$ & 
$\lambda_E $ &=& $ \phantom{-}0.8835616$ \\ 
$\lambda_D $ &=& $ -1.8024639$ & 
$\lambda_Q $ &=& $ -0.2965286$ \\ 
$\lambda_3 $ &=& $ -0.02561451$ & 
$\lambda_2 $ &=& $ -0.8882496$ \\ 
$\lambda_{RL} $ &=& $ -1.021554$ & 
$\lambda_{LR} $ &=& $ \phantom{-}0.0014201883$ \\ 
$v_\varphi $ &=& $ \phantom{-}1717.639$ & 
$v_\phi $ &=& $ \phantom{-}1738.4050$ \\ 
$g' $ &=& $ -0.9171299$ \\\bottomrule 
    \end{tabular}
  \end{center}
  \caption{Best fit point to the data for case (II.).
  We do not list $M_R$, $\lambda_{\nu 1,2}$ and $\lambda_N$ because their precise values do not affect the results.}
  \label{tab:bestfitII}
\end{table}

Below we show the coupling matrices of $Z'$ in the mass basis for our best fit point (I.). The analogous
couplings for point (II.) are qualitatively the same.
\begin{equation}  \label{eq:gLL}
  \hat g^\ell_{L} =
	\left(
\begin{array}{cccc}
 -0.2669 & 0.2798 & 0.01800 & 0.3425 \\
 0.2798 & -0.2934 & -0.01890 & -0.3591 \\
 0.01800 & -0.01890 & 0.9996 & -0.0100 \\
 0.3425 & -0.3591 & -0.0100 & -0.4393 \\
\end{array}
\right)
\end{equation}
\begin{equation}  \label{eq:gLR}
  \hat{g}^\ell_{R} =
	\left(
\begin{array}{cccc}
 -0.2674 & -0.2799 & -0.01800 & -0.3427 \\
 -0.2799 & -0.2930 & -0.01890 & -0.3588 \\
 -0.01800 & -0.01890 & 0.9996 & -0.0100 \\
 -0.3427 & -0.3588 & -0.0100 & -0.4392 \\
\end{array}
\right)
\end{equation}
\begin{equation}  \label{eq:gDL}
  \hat{g}^d_{L} =
   \left(
\begin{array}{cccc}
 -0.000100 & -0.004400 & 0.0007000 & 0.01030 \\
 -0.004400 & -0.1539 & 0.02570 & 0.3613 \\
 0.0007000 & 0.02570 & 0.9994 & -0.006300 \\
 0.01030 & 0.3613 & -0.006300 & -0.8454 \\
\end{array}
\right)
\end{equation}
\begin{equation}\label{eq:gDR}
  \hat{g}^d_{R} =
  \left(
\begin{array}{cccc}
 -0.8418 & 0.05670 & 0.01670 & 0.3604 \\
 0.05670 & -0.003800 & -0.001100 & -0.02430 \\
 0.01670 & -0.001100 & 0.9998 & 5.546\times10^{-6} \\
 0.3604 & -0.02430 & 5.546\times10^{-6} & -0.1542 \\
\end{array}
\right)
\end{equation}


\section*{ACKNOWLEDGMENTS}
A.T.\ is grateful to J.\ Heeck, M.\ E.\ Krauss, and A.\ Vicente for useful discussions.
S.R.\ acknowledges that this material is based upon work supported by the Department of Energy under Award Number DE-SC0011726.
This work has partially been supported by the German Science Foundation (DFG) within the SFB-Transregio TR33 ``The Dark Universe".
S.R.\ acknowledges partial support from the Alexander von Humboldt Foundation and the Bethe Center of Bonn University where this work was begun, 
and the Kobayashi-Maskawa Institute, Nagoya University where this work was completed.


\clearpage
\newpage
\bibliographystyle{utphys}
\bibliography{bibliography}

\end{document}